\documentclass[11pt,aps,nofootinbib,prd,aps,epsf,floats,axodraw,amsmath,amssymb,amsfonts]{revtex4}
\usepackage{amsmath, amssymb}
\newcommand{\mathsym}[1]{{}}

\usepackage{graphicx}
\usepackage{amsmath}
\usepackage{amssymb}
\usepackage{bm}
\newcommand{\be}{\begin{equation}}
\newcommand{\ee}{\end{equation}}
\newcommand{\bea}{\begin{eqnarray}}
\newcommand{\eea}{\end{eqnarray}}

\newcommand{\rem}[1]{}
\newsavebox{\PSLASH}
 \sbox{\PSLASH}{$p$\hspace{-1.8mm}/}
 
\renewcommand{\theequation}{\thesection.\arabic{equation}}
\newcounter{saveeqn}
\newcommand{\add}{\addtocounter{equation}{1}}
\newcommand{\alpheqn}{\setcounter{saveeqn}{\value{equation}}%
\setcounter{equation}{0}%
\renewcommand{\theequation}{\mbox{\thesection.\arabic{saveeqn}{\alph{equation}}}}}
\newcommand{\reseteqn}{\setcounter{equation}{\value{saveeqn}}%
\renewcommand{\theequation}{\thesection.\arabic{equation}}}

 \newsavebox{\notrightarrow}
 \sbox{\notrightarrow}{$\to$\hspace{-4mm}/}
 
 \newsavebox{\PARTIALSLASH}
 \sbox{\PARTIALSLASH}{$\partial$\hspace{-1.6mm}/}
 
 \newsavebox{\ASLASH}
 \sbox{\ASLASH}{$A$\hspace{-2.1mm}/}
 
 \newsavebox{\KSLASH}
 \sbox{\KSLASH}{$k$\hspace{-1.8mm}/}
 
 \newsavebox{\LSLASH}
 \sbox{\LSLASH}{$\ell$\hspace{-1.8mm}/}
 
 \newsavebox{\QSLASH}
 \sbox{\QSLASH}{$q$\hspace{-1.8mm}/}
 
 \newsavebox{\DSLASH}
 \sbox{\DSLASH}{$D$\hspace{-2.2mm}/}
 
 \newsavebox{\DbfSLASH}
 \sbox{\DbfSLASH}{${\mathbf D}$\hspace{-2.8mm}/}
 
 \newsavebox{\DELVECRIGHT}
 \sbox{\DELVECRIGHT}{$\stackrel{\rightarrow}{\partial}$}
 
 \newcommand{\blue}{\IfColor{\textCadetBlue}{}}
\newcommand{\black}{\IfColor{\textBlack}{}}
\newcommand{\red}{\IfColor{\textRed}{}}
\newcommand{\green}{\IfColor{\textOliveGreen}{}}
\newcommand{\lila}{\IfColor{\textRedViolet}{}}

\begin{document}
\begin{flushright}
 [math-ph]
\end{flushright}
\title{Many-Worlds Interpretation of Quantum Mechanics:\\
A Paradoxical Picture}

\author{Amir Abbass Varshovi}\email{ab.varshovi@sci.ui.ac.ir/amirabbassv@gmail.com/amirabbassv@ipm.ir}

\affiliation{Faculty of Mathematics and Statistics, Department of Applied Mathematics and Computer Science, University of Isfahan, Isfahan, IRAN.\\
   School of Mathematics, Institute for Research in Fundamental Sciences (IPM), P.O. Box: 19395-5746, Tehran, IRAN.}
\begin{abstract}
   \textbf{Abstract\textbf{:}} The many-worlds interpretation (MWI) of quantum mechanics is studied from an unprecedented ontological perspective based on the reality of (semi-) deterministic parallel worlds in the interpretation. It is demonstrated that with thanks to the uncertainty principle there would be no consistent way to specify the correct ontology of the Universe, hence the MWI is subject to an inherent contradiction which claims that the world where we live in is unreal. \\
       
\noindent \textbf{Keywords\textbf{:}} Many-Worlds Interpretation of Quantum Mechanics, (Semi-) Deterministic World, Fiducial World, Inclusive Consciousness.
\end{abstract}

 \maketitle


\section{Introduction}
\label{introduction}

\par About a century after its inception, quantum mechanics is still in a perplexity to provide a precise comprehensive interpretation of its peculiar features. Even though, the orthodox interpretation of quantum theory, the so called Copenhagen interpretation, soon gained a worldwide acceptance for describing the quantum behaviors of the objective world around us, this quandary is still felt in the basics of quantum mechanics \cite{cabello}. The most profound consideration to this dilemma dates back to 1930s when the Einstein-Podolsky-Rosen criticism of the Copenhagen interpretation of the theory, the EPR paradox \cite{epr}, revealed an unexpected aspect of quantum physics which violates the main principle of special relativity allowing information to be transmitted faster than light. This phenomenon that was the most serious objection to the interpretation until then, was referred to as the \emph{entanglement} effect by Schrodinger \cite{sch3}.

\par However, despite to the responses and the paraphrases of the founding fathers of quantum mechanics \cite{bohr, sch1, sch2}, the EPR paradox led to significant implications to the interpretation and yield the birth of other intellectual paradoxes in next years most of which were fundamentally rooted in the elaborately planned Schrodinger's cat experiment and its ontological expression of reality.\footnote{This experiment is an attempt to explain the reality of the counterintuitive nature of quantum mechanics by transferring the superposition states to macroscopic objects \cite{sch3}. Here we should emphasize that through this article by \emph{reality} we mean the \emph{observed reality} defined based on the practical features of the pragmatic worldview in the literature \cite{epr, heisenberg-ph}. Thus, hereby we hinder the subjective speculations with unobserved effects, including the hidden variables and the QBism interpretation of quantum mechanics. We acknowledge that via the Copenhagen interpretation the observable reality undergoes an ontological revision in favor of the Heisenberg's uncertainty principle and its inevitable consequence, i.e. the non-locality. (For a more detailed discussion on ontological and epistemological perspectives of reality and the relevant semantic disputes see also \cite{penrose2, dewitt}.)} Among them, as an immediate generalization of the Schrodinger's cat paradox to the consciousness problem, the Wigner's friend experiment \cite{wigner, deutsch}, has been considered delicately in some extended scenarios leading to critical discoveries of serious intellectual inconsistencies in the Copenhagen interpretation \cite{fr0, sudbery, fr, brukner1, brukner2, laloe, healey1, healey2, dieks, lombardi}.\footnote{Even, revisiting the Schrodinger's cat experiment itself, in some definite viewpoints, has yield some paradoxical aspects of the Copenhagen interpretation such as the problem of quantum suicide and immortality \cite{lewis0, lewis}. See also \cite{hobson} for more detailed discussions/suggestions and \cite{castel} for fatal features of the experiment.} \footnote{Other intellectual inconsistencies of the Copenhagen interpretation are listed in diverse realms of quantum physics such as elementary particle physics (the Hardy's paradox) \cite{hardy1, hardy2, hardy3, hardy4}, quantum cosmology \cite{cos1, cos2} (see also \cite{shestakova} for a counterproposal viewpoint), the Hawking's paradox \cite{hawking, susskind} and other inconsistencies in quantum viewpoints to general relativity \cite{pavsic, time, camacho} (besides the well-known contradictions of quantum gravity). For a more complete list of such paradoxes and their explanations see \cite{smolin, qmparadox}.}

\par Although the subsequent experiments confirmed the completeness of the Copenhagen interpretation despite the EPR paradox claim \cite{aspe1, aspe2}, recent developments of experimental physics have established some inconsistencies of the interpretation by running quite complex empirical tests in last two decades. As one of the pioneers, the Afshar's experiment \cite{afshar1, afshar2} reported a controversial violation of the wave-particle duality in the early years of the present century. A few years lated weak measurements \cite{hardy-exp1, hardy-exp2, hardy-exp3, hardy-exp4} also revealed objective contradictions in particle-anti-particle annihilation process due to the Hardy's paradox \cite{hardy1, hardy2, hardy3, hardy4}. Even more recently, some spectacular experiments, fundamentally designed based on the extended versions of the Wigner's friend scenario, released fatal conflicts in basic foundations of the Copenhagen interpretation  \cite{proietti, kok}. All in all, these facts prove the obligation of a basic reconsideration to the Copenhagen interpretation of quantum mechanics.

\par Actually, after releasing the EPR paradox, the Copenhagen interpretation as the only reliable acknowledged expression of quantum mechanics in its time, was gradually distrusted in next years and encountered with diverse rival interpretations which were proposed to explain the deceptive effects of the theory from distinct outlooks and philosophies. Among them, the Many-Worlds interpretation (MWI) of quantum mechanics, proposed by Hugh Everett in 1950s \cite{everett1, everett2},\footnote{And was popularized by Bryce DeWitt during 1970s \cite{everett4, dewitt-mwi}.} has widely been considered as the most reliable alternative for the Copenhagen interpretation \cite{everett3, tegmark}. Unlike the Copenhagen interpretation which embraces a single world and its observable reality, the MWI construes the Universe as an expanding cluster of infinitely many increasingly divergent parallel worlds. It is, in principle, a deterministic interpretation which includes the non-locality by introducing parallel worlds to realize or to embody each possible history of outcomes of quantum mechanical measurements \cite{everett1}.\footnote{In this sense, the \emph{Universe} is sometimes referred to as the \emph{Multiverse}. Also what we call \emph{world} here is seldom addressed to as \emph{universe}. Actually, through this paper we follow the terminology of \cite{vaidman}.} In particular, the status of the wavefunction for the MWI is ontological in contrast to that in the Copenhagen interpretation as epistemic, summarizing information about the results of measurements \cite{becker}.

\par Actually, in the MWI, despite of the Copenhagen interpretation, the central role of observer in the process of quantum measurement is disclaimed by replacing wavefunction collapse with quantum decoherence. On the other hand, the observer's role lies at the central part of all paradoxes listed above. Hence, the MWI is considered as one of the most plausible adequate alternatives to resolve the inconsistencies of the Copenhagen interpretation such as those of the EPR paradox, the Schrödinger's cat and the Wigner's friend thought experiments,\footnote{Including their extended scenarios, e.g. the Frauchiger-Renner's discussion \cite{fr0, fr}. See also the Brukner's \cite{brukner1, brukner2} and the Healey's \cite{healey1} expression of the experiment.} the Hardy's paradox and even the wave-particle duality.\footnote{See \cite{mwi-epr} for the EPR paradox. Also, see \cite{mwi-oxford} for more discussions about the Schrodinger's cat and the Wigner's friend experiments..}

\par However, despite of the proposals that put forward to assess the practical accuracy of the MWI \cite{test1, test2, test3}, it does not seem to be testable with current and any foreseeable future experimental capability \cite{deutsch, vaidman}. In fact, all such tests rely on contrived conditions and idealized circumstances which would be infeasible or exceedingly difficult to realize in the real world. Also, as the most interesting candidate scenario to distinguish the MWI from all single world-base interpretations of quantum mechanics, the quantum suicide gedankenexperiment, would not also work in the real world \cite{suicide}. Therefore, the theoretical premises of testability of the MWI are controversial even among the proponents of the interpretation \cite{dewitt}. Although it will threaten the refutability of the interpretation,\footnote{But, however, there are other claims and counterproposals against this failure \cite{vaidman}.} it is also seen as an opportunity for the MWI not to be criticized or falsified practically.

\par On the other hand, in contrast to empirical confusion about the MWI, it has already faced to a number of intellectual criticisms including the philosophical objection due to the law of parsimony,\footnote{Also known as the Occam's razor. See \cite{vaidman} for more discussions.}  the preferred basis ambiguity,\footnote{See \cite{wallace2010} for the critique and its solution.} the ontological insufficiency of the wavefunction of the Universe,\footnote{See \cite{bell, allorri} for the detail of the objection and \cite{vaidman} for the reply.} and the incoherence probability problem.\footnote{See \cite{wallace2003} for the analysis of the problem and its solution.} However, it is purported by the supporters of the MWI that all such critiques stem in incomplete viewpoints and misunderstandings \cite{vaidman}. Hence, it seems that the MWI can keep developing increasingly without any serious experimental/intellectual obstacle. But, actually, in contrast to such privileges, the MWI is still far from being a widely accepted interpretation just like the Copenhagen interpretation is, since there would be no easily reachable consensus on its ontological perspective based on introducing infinitely many worlds that we do not see.

\par To overcome the dilemma arisen in understanding quantum mechanics via the unprovable MWI, it is mandatory to figure out all its possible aspects with a careful scrutiny of the very assumptions and a comprehensive contemplation of the ontological results. Only then after we can argue and judge about the possibility of replacing the Copenhagen interpretation with the MWI for explaining quantum physics. It is the main objective of this paper to make some troubling features of the MWI more apparent and to shed more light on its reliability. Actually, through this research we will report an inherent intellectual paradox in the very ontological foundations of the MWI by enquiring: \emph{Who can judge about the reality of other parallel worlds?} But here, the interrogative pronoun "\emph{Who}" refers to sentient beings of all parallel worlds introduced in the MWI and the whole question enquires: \emph{Whose interpretation is correct?} Despite of its philosophical statement we try to give a mathematical response to the question. We will discover a fundamental conflict among the ontological assumptions of the MWI leading to unreality of the world where we live in. 


\par
\section{Many-Worlds Interpretation: Illustrative or Seductive}
\setcounter{equation}{0}

\par Despite its abilities to describe a significant portion of quantum mechanical results, the MWI is substantially a paradoxical approach. In other words, although the MWI is the second most popular and thought-provoking way to interpret quantum mechanics \cite{tegmark}, a careful investigation will find it to include a contradiction in its basic assumptions. In fact, this inconsistency is rooted in ascribing a real world to each possible history of measurement outcomes. More precisely, this assumption leads to a serious problem for defining the reference worlds whose sentient beings are eligible and capable to judge about the ontological aspects of quantum physics. Since, in principle, assuming the MWI there would be some parallel worlds whose inhabitants will totally oppose to register the MWI for describing the Universe. This, in principle, shows that the only way to accept the MWI is to rely on the reference worlds. But, it is established here that assuming a reference world (or a group of reference worlds) is theoretically impossible, hence, the MWI is a paradoxical picture.

\par Therefore, based on its very assumptions, the MWI is not rationally acceptable, unless it undergoes a fundamental revision within a firm framework. Such a framework, as we see below, should include an overall assessment of the Universe as the ensemble of all parallel worlds in the MWI and their embraced registered established theories of physics. This, in its turn may cause a statistical/stochastic investigation on various theories of physics, with almost no limit for exotic laws of nature, to study the aspects of reality and the promised comprehensive interpretation of quantum physics. However, this extreme complexity shows that the MWI will have a long way to go before it can be accepted theoretically.

\par To prove the claim mentioned above with details we have to consider two following essential enquiries:\\

\par \textbf{Q1-} \emph{Assume that the MWI is correct. Are there some parallel worlds whose agents reject the existence of some other parallel worlds based on the history of their measurement outcomes?}\\
\par \textbf{Q2-} \emph{Assume that the MWI is correct. Are the whole ontologies of the possible universes described by the sentient beings of all parallel worlds (based on their observed reality) equivalent in every way?}\footnote{Actually, the \emph{possible universe} for any agent is the largest cluster of parallel worlds each of which is feasible based on the laws of nature discovered from the history of the agent's measurement outcomes. For instance, for us, the possible universe is equivalent to the \emph{Universe} as a whole, just as we assume in the MWI. On the other hand, our possible universe excludes the parallel worlds with no experience of Pauli's exclusion principle. Upon the laws of our theory of physics, such worlds are not real and do not belong to the Universe.}\\

\par We will prove that the enquiries \textbf{Q1} and \textbf{Q2} both do have positive responses, leading us to the following paradoxical conclusion:\\

\par \textbf{Conclusion:} \emph{Assuming the MWI as a correct theory, there would be a parallel world whose agents acquire a correct complete description of the possible universe and reject some parts of the ontological claims of the MWI. Hence, some parallel worlds assumed in the MWI, including ours, are both real (by our ontology) and unreal (based on the ontology of the agents) simultaneously!}\\

\par More precisely, the above conclusion asserts that the ontology of the MWI rejects itself, hence an inherent conflict. Our arguments to prove the mentioned conclusion are as following:\\

\par \textbf{Q1:} This inquiry can be rewritten within the following wording:\\

\par \emph{Assuming the MWI as a correct theory, is there any agent in some parallel world whose description of the possible universe (based on the reality observed in her/his world) disclaims the existence of some other parallel worlds and consequently rejects the ontology of the MWI?}\\

\par \textbf{Reply: YES!}\\

\par
\begin{figure}
 \includegraphics[width=\linewidth]{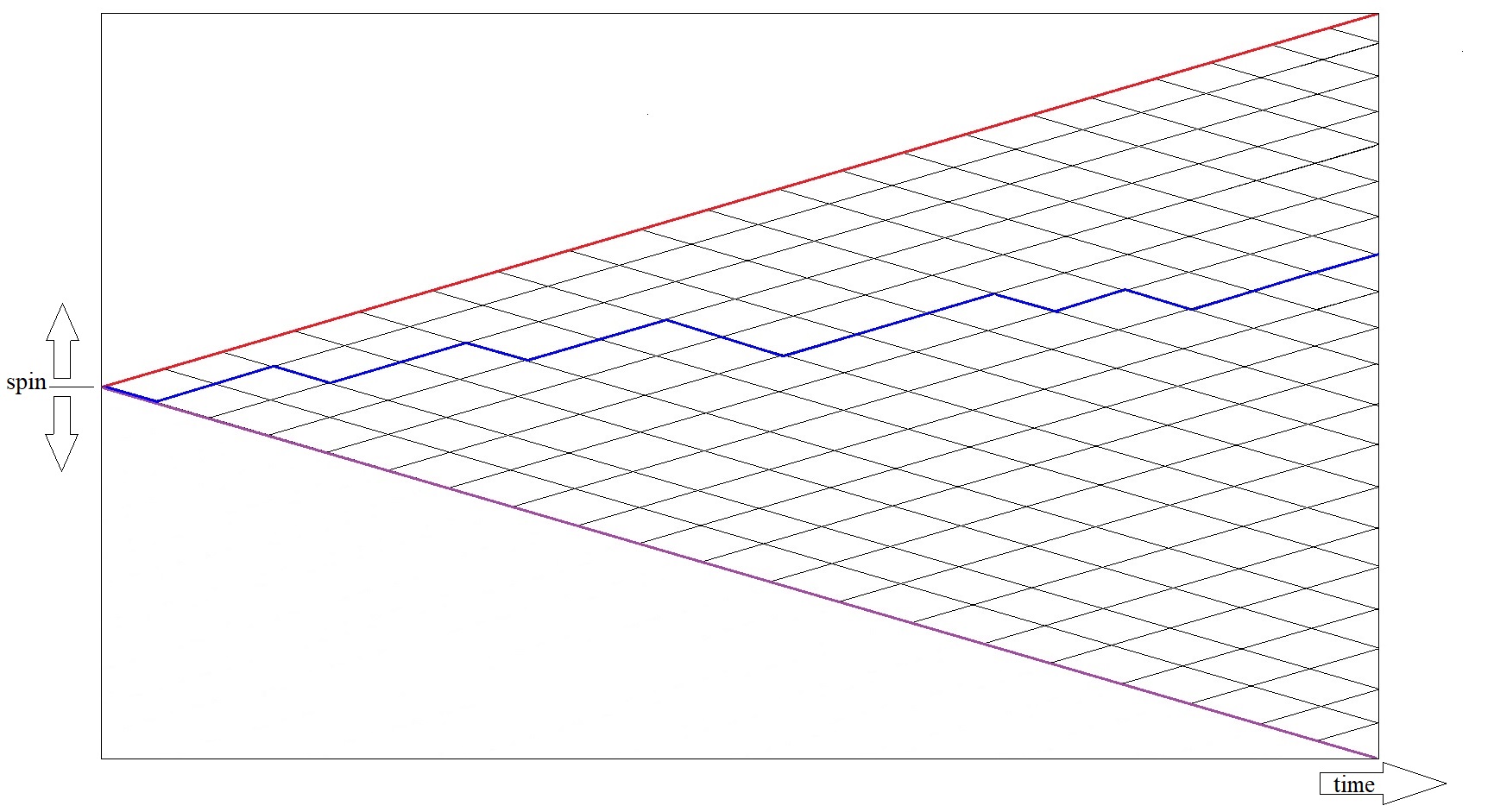}
\caption{\textbf{The Many-Worlds Interpretation of the Universe:} A schematic diagram for the MWI of quantum mechanics of electron spins (measured on the $z$-axis) and the corresponding Universe. The simplest example of the MWI generated by a sequence of measurements on the superposition state $\left| \psi \right\rangle =\frac{1}{\sqrt{2}} \left| \uparrow \right\rangle + \frac{1}{\sqrt{2}} \left| \downarrow \right\rangle$. The horizontal axis indicates the direction of time and the vertical axis exhibits the amount of the measured electron spin. Any single vertex displays a measurement in which the outcomes are drawn as two segments: Each segment with positive slope correspond to observing a $+\frac{1}{2}\hbar$ spin ($\left| \uparrow \right\rangle$) and that with negative slopes is related to $-\frac{1}{2}\hbar$ one ($\left| \downarrow \right\rangle$). Hence, any possible right-going zigzag path depicts a parallel world in the MWI. The straight paths seen at the extreme edges of the lattice (the red line and the purple line) are actually illustrating two (semi-) deterministic worlds with the only $+\frac{1}{2}\hbar$-spin (red) or $-\frac{1}{2}\hbar$-spin (purple) electrons. The blue path also shows a non-deterministic world with a different theory of quantum mechanics in which the possibility of observing $+\frac{1}{2}$-spin electrons is twice to that of $-\frac{1}{2}$-spin ones.}
\label{fig1}
\end{figure}

 \textbf{Reasoning:} Actually, one can easily build up real worlds, based on the MWI ontology, which reject some acknowledged quantum mechanical aspects (principles) and embrace an entirely different theory of physics. In fact, such peculiar physics may include semi-deterministic theories,\footnote{By semi-deterministic theories or semi-deterministic worlds we mean the parallel worlds in which some sort of quantum mechanical measurements (specially, those which are implemented in laboratories and preferably have no immediate results on biological status of the world, if there is one) gain absolutely predicted outcomes. For example, we can assume the world in which all measurements of the electron spin of the superposition states lead to a single result, but the rest experiments of quantum mechanics retain the usual probabilistic features.} deterministic theories,\footnote{By deterministic theories or deterministic worlds we mean the parallel worlds in which all quantum mechanical tests lead to definite predicted results (from a moment onward and for all superposition states). In fact, a deterministic world is defined by employing the inductive approach to semi-deterministic worlds.} deformed quantum theories,\footnote{By deformed quantum theories or deformed quantum worlds we mean the parallel worlds with different probabilities of observations. For example, we may consider a world in which the probability of observing $+\frac{1}{2}\hbar$-spin electron, i.e. $\left| \uparrow \right\rangle$, is not equal to $|a|^2$ in the superposition state $\left| \psi \right\rangle = a \left|  \uparrow \right\rangle + b \left| \downarrow \right\rangle$, for each $a,b \in \mathbb{C}$ with $|a|^2+|b|^2=1$. This quantum theory, although adheres to the uncertainty principle and may be described with (a likely modified version of) the Schrodinger's equation, should be equipped with a distinct Hilbert space and its corresponding mathematics for electrons. Also, in (FIG. \ref{fig1}) a deformed quantum theory (world) is depicted wherein the probability of observing an upward-spin electron $\left| \uparrow \right\rangle$ (with $+\frac{1}{2}\hbar$ spin on the $z$-axis) is twice to that of the downward-spin electron $\left| \downarrow \right\rangle$ (with $-\frac{1}{2}\hbar$ spin, measured on the same axis) in the superposition state $\left| \psi \right\rangle = \frac{1}{\sqrt{2}} \left|  \uparrow \right\rangle + \frac{1}{\sqrt{2}} \left| \downarrow \right\rangle$ (the blue path in (FIG. \ref{fig1})). However, despite to the different formulation of such a theory, with thanks to the uncertainty principle, this world may embrace the ontology of the MWI.} and in general manually adjusted laws.\footnote{By manually adjusted laws or manually adjusted worlds we mean parallel worlds in the MWI wherein the probabilities of measurement outcomes have been adjusted manually to capture (almost) controlled laws of nature. For example, we can assume a real world in which the electron spin of any superposition state is $+\frac{1}{2}\hbar$ during autumn and winter and $-\frac{1}{2}\hbar$ in spring and summer.} All in all, according to the MWI there do exist real parallel worlds in which some of the objective realities of ours\footnote{The possessive pronoun "ours" (just like "our" and "us") is used in this article to address to the author's and the readers' "world" and "nature", until the moment we can think and speculate about the MWI.} are concealed or eliminated from the cognition of their sentient beings. Therefore, even if in such a parallel world there are evidences for existence of other parallel worlds and the possible universe, its agents' interpretation of the corresponding ontology will not be compatible to that of ours.

\par Specifically, these incompatibilities occur in (semi-) deterministic worlds. As described above, such worlds can be designed by arranging the recorded history of outcomes of measurements to gain some controlled laws of nature with certain predictability. An example of a semi-deterministic world\footnote{Actually, any parallel world which (from a definite moment onward) recognizes a clear ordering in some sort of quantum mechanical tests (e.g. measurement of the spin of electrons) is semi-deterministic. More precisely, a semi-deterministic world is a parallel world which violates the Heisenberg's uncertainty principle for one or more quantum observables.} is the \emph{world of upward-spin electrons} wherein all superposition states of electron are measured as $+\frac{1}{2}\hbar$-spin in experiments.\footnote{Through this paper, all spin measurements are assumed to be performed on the $z$-axis.} Practically, there would be no superposition amongst \emph{up} and \emph{down} spin states for electrons,\footnote{Obviously, assuming such a world is permitted in the MWI. In particular, by the uncertainty principle the state vector of any electron is actually (almost always) a non-trivial superposition of $+\frac{1}{2}\hbar$-spin and $-\frac{1}{2}\hbar$-spin states. Thus, it can be claimed that there would be no evidence for (non-entangled) electrons of $-\frac{1}{2}\hbar$-spin in this world. (However, since the probability of finding a pure state of $-\frac{1}{2}\hbar$-spin electron in nature is ignorable, it would be neglected statistically and be reported as an experimental error whenever observed.)} and thus the spinor structure of electron (i.e. the existence and the reality of electrons of $-\frac{1}{2}\hbar$ spin) is ruled out in action (see (FIG. \ref{fig1})).

\par More precisely, according to the MWI assumptions we can find real (semi-) deterministic worlds with different objective features of reality, hence with restricted deformed ontology of the possible universe.\footnote{Even though this can endanger the existence of any inhabitant sentient beings and generally of life (as we acknowledge it) in such worlds, it is an inevitable consequence of the MWI of quantum mechanics.} For instance, the sentient beings of the world of upward-spin electrons will not even think about the existence of parallel worlds with downward electron spins, just like we do not consider a parallel world of electrons with negative mass.\footnote{We should note that assuming the existence of such exotic worlds is out of the fundamentals of the MWI, wherein the splitting of parallel worlds take place for all possible outcomes of measurements, as we verify.} Hence, assuming MWI, there would be parallel worlds (e.g. (semi-) deterministic worlds) with exotic interpretations of the possible universe (based on their cognition and upon their ontological perspective) which decline the ontology of the MWI. \textbf{Q.E.D.} \\

\par Let us now discuss the next enquiry:\\

\par \textbf{Q2:} This question can be recounted as the following restatement:\\

\emph{Assuming the MWI as a correct theory, is there any obstruction to define a (practical or semantic) dissociation method on parallel worlds of the MWI (and their inhabitant agents) to recognize the most comprehensive ontology and its possible universe?}\\

\par \textbf{Reply: YES!}\\

\par \textbf{Reasoning:} In fact, we have to prove that there would be no well-defined way to specify some definite fiducial worlds whose sentient inhabitants are the only agents who are eligible for describing the Universe. This assertion is the most fatal consequence of the assumptions of the MWI. More precisely, the main difficulty with the MWI emerges just when one tries to compare the awareness and the ontology of the inhabitants of faithful worlds (i.e. the worlds in which the history of all quantum mechanical tests confirm the Heisenberg's uncertainty principle\footnote{One should note that this definition also contains some of the worlds with deformed versions of quantum mechanics. More precisely, a faithful world is a parallel world which its agents may embrace the MWI entirely, just as e.g. Hugh Everett do in our world.}) and of the agents who are resident in other parallel worlds. Actually, since according to the assumptions of the MWI none of the parallel worlds is substantially the superior, then there would be no predetermined distinguishing criteria to prefer and to admit quantum mechanical description of the possible universe (the MWI) and to discard that of the (semi-) determinist agents.\footnote{Perhaps one may speculate about the \emph{measure of existence} as an appropriate criteria to recognize the fiducial worlds \cite{vaidman, everett2, test2} by assuming; \emph{A fiducial world would have the most measure of existence}. But actually, this definition will not work properly, since there are (semi-) deterministic worlds which their measure of existence is equal to (or even is larger than) that of a faithful world. For example, all parallel worlds in (FIG. \ref{fig1}) are of the same measure of existence. Moreover, if we replace $\left| \psi \right\rangle=\frac{1}{\sqrt{2}}\left| \uparrow \right\rangle+\frac{1}{\sqrt{2}} \left| \downarrow \right\rangle$ with $\left| \psi \right\rangle=\sqrt{\frac{2}{3}}\left| \uparrow \right\rangle+\frac{1}{\sqrt{3}} \left| \downarrow \right\rangle$ in (FIG. \ref{fig1}), then the (semi-) deterministic world depicted with the red line at the top of the figure would have an absolutely larger measure of existence rather than any faithful world.}

\par In other words, upon its very assumptions, the MWI is not capable to distinguish fiducial references amongst its claimed parallel worlds and hence is not able to recognize the correct interpretation of the possible universe. That is assuming any preference or priority for specifying some parallel worlds would be an unprecedented hypothesis. Here, we show that this assumption is also inherently inconsistent provided the MWI be correct. Actually, the machinery we introduce here can be applied to any other concept/characteristic which is defined upon the history of measurement outcomes for separating parallel worlds. More precisely, our analysis here demonstrates that the parallel worlds can never be distinguished by any statistical subjective/objective characteristic.

\par At the first glance, one should note that ascribing any separating characteristic to a group of specific real worlds and their inhabitants, is theoretically equivalent to append a consciousness-like attribute to the sentient beings who accomplish measurements in the Winger-von Neumann interpretation of wave collapse \cite{wigner,vonneumann}. Just as consciousness separates the observer from the cat in the Schrodinger's cat experiment, the inclusive consciousness, if exists, must separate the fiducial worlds from the other ones in the MWI. Let us refer to this characteristic of the fiducial worlds, as the \emph{inclusive consciousness}. Obviously, since the MWI is assumed to be a correct theory, the fiducial worlds who enjoy this property the most must respect the uncertainty principle, hence being the faithful worlds.

\par Actually, as we show in the following, the inclusive consciousness is a superfluous concept with no specific efficacy. To see this clearly we first show that the inclusive consciousness has not to be specified to some definite individuals, but it must be continuously distributed over all parallel worlds. Actually, this is the immediate consequence of physical laws as statistical consequences. Thereafter, we will mathematically prove that the distribution is inevitably a constant function, which in its turn, leads to uselessness of the concept, indicating all parallel worlds have the same amount of inclusive consciousness and their sentient beings acquire enough eligibility to describe the ontology of the possible universe. In fact, the monotony of the distribution is inferred by considering the Universe of (FIG. \ref{fig1}) (as a simple example of the MWI) and stems in the fact that both the set of faithful worlds (fiducial worlds) and that of the (semi-) deterministic worlds are dense in the Universe. \\

\par In the following we prove the above claims accordingly: \\

\par \textbf{1- Inclusive Consciousness is a Continuous Distribution}\\

\par To approve that the inclusive consciousness is a continuous distribution over the Universe, one should first note that a fiducial world is a world whose sentient inhabitants confirm the uncertainty principle of quantum mechanics based on the recorded statistics (history) of all quantum mechanical measurement outcomes. On the other hand, based on its statistical essence, the uncertainty principle needs enough time for performing sufficient numbers of experiments to be registered or dismissed in a parallel world.\footnote{We should note that this is a theoretical fact which does not depend on the extent of inhabitants cognition and ingenuity.} Then, the inclusive consciousness, arises or disappears over time. Consequently, a semi-deterministic world, separated from a faithful world (a fiducial world) at some specific branching point, also carries inclusive consciousness and there would be a no definite time for losing it thoroughly. Hence, a (semi-) deterministic world have some inclusive consciousness depending on how much it is close (in some sense) to a faithful world.

\par In particular, observing the only upward spins in a (semi-) deterministic world could be statistically interpreted as a finite portion of a longer sequence of observations which confirm the uncertainty principle as a whole. In other words, at each finite specific moment none can talk about absolute deterministic worlds in the Universe. The only thing we can probably comment on is how much a world is statistically inclined to be (semi-) deterministic. We assume that the more of this tendency a parallel world may have (upon the history of measurement outcomes), the less inclusive consciousness it would include. Actually, according to the statistical essence of the uncertainty principle this is the only possible way to define inclusive consciousness in a consistent disposition. Consequently, to introduce inclusive consciousness correctly one has no choice except employing a continuous distribution over the parallel worlds. \\

\par \textbf{2- Inclusive Consciousness is a Constant Function}\\

\par Let us redefine that the concept of inclusive consciousness so that it could be evaluated with real numbers\footnote{The formulation which we workout here is too flexible to be generalized to any topological set such as $\mathbb{C}$, $\mathbb{H}$, and $\mathbb{R}^n$, $n>1$.} assigned to parallel worlds to determine the credit rate or the amount of validity of their inhabitants' understanding of the uncertainty principle and the MWI.\footnote{We should note that this is not a psychological assessment, but an evaluation based on the statistics of the history of the outcomes of quantum mechanical tests.} Therefore, considering $U$ as the set of all parallel worlds, the inclusive consciousness is evaluated with a continuous function, say $f:U \to \mathbb{R}$, which takes the same value for faithful worlds. Moreover, for parallel worlds $w_1$ and $w_2$, we must have $f(w_2) > f(w_1)$ if and only if the ontology of the inhabitants of $w_2$ is more comprehensive and more accurate with respect to that of the residents of $w_1$. Thus, we may assume $f(w_2) > f(w_1)$ whenever $U_1 \subset U_2 \subseteq U$, wherein $U_1$ (resp. $U_2$) is the possible universe of the agents of $w_1$ (resp. $w_2$).\footnote{ For example, the world of upward-spin electrons (which its sentient beings agree with us on all quantum concepts except the spin of electrons), as a semi-deterministic world, encompasses more inclusive consciousness with respect to a deterministic world (which its inhabitants do not realize any evidence to figure out and to trust in quantum mechanics and the uncertainty principle), and we hope that $f$ could recognize this dissimilitude.} For instance, if $w_2$ is a faithful world and $w_1$ is a semi-deterministic world we should find:\footnote{Similarly, if $w_2$ is a semi-deterministic world and $w_1$ is a deterministic world once again we obtain $f(w_2)>f(w_1)$.}
\begin{equation} \label{1}
 f(w_2)>f(w_1).
 \end{equation}

\par In the following we show that $f$ violates (\ref{1}) for the Universe depicted in (FIG. \ref{fig1}). In order to do this, we must first study $U$ of (FIG. \ref{fig1}) with some more details. Actually, here $U$ can be considered as the Universe generated by an agent (and its copies in parallel worlds) during the process of measuring the spins of elements of an ensemble which includes infinitely many copies of the superposition state $\left| \psi \right\rangle = \frac{1}{\sqrt{2}}\left| \uparrow  \right\rangle +\frac{1}{\sqrt{2}} \left| \downarrow \right\rangle$ of electron.

\par Initially, we claim that there exists an isomorphism between $U$ and $[0,1]$. To see this let us assume that the ensemble elements are labeled with natural numbers $\mathbb{N}$. Then, if we assign a $1$ to each $+\frac{1}{2}\hbar$ outcome, and a $0$ to any observed $-\frac{1}{2}\hbar$, each parallel world is uniquely shown with a sequence of integers $1$ and $0$. For instance, the blue path (world) in (FIG. \ref{fig1}) depicts $01101101100111010111\cdots$. Let us assume these sequences as decimal digits in base two, e.g. the above blue world be denoted as $0.01101101100111010111\cdots$ in base two. Actually, this will assign a unique real number in $[0,1]$ to each parallel world.\footnote{We note that $0=0.000\cdots$ and $1=0.111\cdots$. In fact, the former is depicted with the purple line and the later one is shown by the red line in (FIG. \ref{fig1}).} Conversely, each number of $[0,1]$ is also indicating a parallel world in $U$.\footnote{It is the immediate consequence of our assumption: the ensemble contains infinitely many elements.} Consequently, one simply finds; $U=[0,1]$. Thus, from now on any parallel world is understood as a real number in $[0,1]$, and the inclusive consciousness is considered to be evaluated with a real function as; $f:[0,1] \to \mathbb{R}$. Although, this redefinition of $f$ may cause its continuity to be lost, but fortunately, it can be shown that $f$ is still a continuous function.

\par To study the continuity of $f$ let us compare the topology of $U$ and $[0,1]$. Note that the closer the two points are to each other in $[0,1]$, their more decimal digits will be equal to. On the other hand, the more equal decimal digits two parallel worlds have, the longer common history (and statistics) of experiments they have passed. Therefore, the standard topology of $[0,1]$ coincides with the overall understanding of the correlations of the parallel worlds: \emph{The distance of two points in $[0,1]$ reveals the disagreement of the cognitions about the objective reality and the disparity of the ontologies of the possible universe the agents of the two corresponding worlds may gain based on their prior experiments}. Hence, the standard topology of $[0,1]$ respects the topological structure of $U$ from the ontological viewpoint. Due to this property, once again we consider $f:[0,1] \to \mathbb{R}$ as a continuous function: \emph{The more compatible ontologies (interpretations of the possible universe) the agents of two worlds may obtain, the closer evaluation of inclusive consciousness they will have}.

\par A (semi-) deterministic world, as we consider here,\footnote{Since we restrict ourselves to investigate the Universe generated by observing the spin of electrons and we ignore all other quantum mechanical observables, the semi-deterministic worlds are in fact deterministic.} is in fact, a parallel world in which the electron spin measurements have absolutely predicted outcomes after some specific moment. In general, from a definite moment onward, observing any perceivable order in the history of the measured spins will result in a deterministic world.\footnote{Actually, if in such a parallel world after a few number of first experiments a definite order is observed in the outcomes of measurements of the spin of electrons, then the uncertainty principle is excluded and we are encountered with a (semi-) deterministic world. As a simple example, if the outcome of even tests is $+\frac{1}{2}\hbar$-spin and the result of odd ones is $-\frac{1}{2}\hbar$ after a specific moment, then we have a (semi-) deterministic world.} This means that any deterministic world is actually shown with some $r\in[0,1]$ which after a number of its first decimal digits, its other decimal digits follow a certain order expressed by repeating a fixed sequence over and over. That is, $r$ has to be a rational number; $r\in \mathbb{Q}$ \cite{hosch}.

\par On the other hand, in faithful worlds, with thanks to the uncertainty principle, there must be no definite order in measurement outcomes. Thus, any faithful world is shown with a real number $s\in[0,1]$ which its decimal digits, will not show any order created by repeating some sequence, hence is an irrational number; $s\in \mathbb{Q}'$. Consequently, upon our definition $f$ must take the same value for irrational numbers of $[0,1]$.

\par Now we are ready to study the analytic structure of $f$. Let us see what we already know about it so far:
\par \textbf{1)} $f:[0,1]\to \mathbb{R}$ is a continuous function.
\par \textbf{2)} $f:[0,1]\to \mathbb{R}$ takes the same value for irrational numbers.\\
\noindent According to two above properties we have no choice but to assume $f$ as a constant function, the fact which violates (\ref{1}). This means that the inclusive consciousness of all (semi-) deterministic words must be equal to that of the faithful ones in the Universe of (FIG. \ref{fig1}). In general, by employing the inductive process for the Universe of several quantum observables we see that no matter how much of them are turned to be predictable, the value of $f$ will still not make a difference.\footnote{The analysis method introduced here can be generalized to any numbers of quantum variables by inductive process. To prove this claim, at the first step we have to show that for any given quantum observable with finitely many possible outcomes the inclusive consciousness is evaluated similarly with a continuous function, i.e. $f:[0,1] \to \mathbb{R}$. Actually, in order to display any parallel world in $U$ with a real number of [0,1] for an observable with $n$ possible outcomes it is sufficient to encode the observed results with integers 0, 1, 2, $\cdots, n-1$, and write down the history of outcomes as decimal numbers in base $n$, just as we did here for the spin of electron. Therefore, $U$ and its topology is thoroughly characterized with $[0,1]$. Moreover, one can simply generalize this formulation for the Universe generated by several numbers of quantum observables. In fact, for $m$ different kinds of quantum mechanical tests the Universe $U$ is encoded with points of $[0,1]^m$. Obviously, hereby any (semi-) deterministic world is shown by a $m$-plet with at least one rational component. Once again we see that both the set of (semi-) deterministic worlds and that of the faithful ones are dense in $U$, and consequently, the inclusive consciousness must also be uniformly distributed over $U$.} That is, $f$ remains constant over all possible parallel worlds in the entire Universe of the MWI. This proves that the inclusive consciousness is an irrelevant null concept for the MWI, just like the ether was included in classical electrodynamics. Hence, assuming the MWI, all agents in all parallel worlds are equally competent from ontological viewpoints. \textbf{Q.E.D.} \\

\par \textbf{Conclusion:} According to the positive responses to \textbf{Q1} and \textbf{Q2}, we conclude that some parallel worlds in the Universe should have paradoxical existence. Since, on the one hand, upon the arguments to \textbf{Q1}, embracing the MWI leads to birth of (semi-) deterministic worlds and their sentient beings interpretations of the possible universe that reject the existence of some parallel worlds with unreal (upon their cognition) features. Specially, the reality of the world which we live in is itself rejected by the ontology of (semi-) determinist agents. On the other hand, due to the argumentations of \textbf{Q2}, assuming the MWI, all interpretations of the possible universe reported by the sentient beings of different parallel worlds, based on their cognition and contemplation about the statistics (history) of quantum mechanical observations, are equally correct from the ontological perspective. Altogether, we conclude that some parallel worlds must and must not exist simultaneously. Surprisingly, the unreality of our world (from the viewpoint of (semi-) determinist agents) is a true fact, just as we recognize it. Hence, the MWI includes paradoxical propositions within its very assumptions.

\par
\section{Summary and Discussion}
\setcounter{equation}{0}

\par In this paper we argued that the MWI is inherently a paradoxical picture because of its exaggerated ontological perspective to the Universe. Actually, considering any possible parallel world to be a real entity would lead to theoretically incontrollable situations such as the violation of the laws of nature as we verify as quantum physics in our world. The most fatal situation arises when we consider the (semi-) deterministic worlds (which statistically reject the Heisenberg's uncertainty principle) as real. Since, it was shown that there would be no well-defined way to discard their agents' ontology as an incorrect picture of the possible universe (even if the measure of existence of such (semi-) deterministic parallel worlds are ignorable). Therefore, there are parallel worlds which are both real (due to our ontology) and unreal (via the ontology of agents in some (demi-) deterministic world) simultaneously. Among them, the most controversial case, is our world which is actually unreal from the ontological perspective of any (semi-) deterministic world.

\par The above argument shows that the MWI is inherently a paradoxical picture of quantum mechanics. To resolve this conflict one may speculate about a number of thought-provoking proposals. As the first solution, one may think about the \emph{chance of detecting (semi-) deterministic worlds} amongst the Universe, i.e. \emph{Since the chance of detecting a (semi-) deterministic world is fairly zero, then we could ignore its agents' ontology}. However, although the chance of finding a (semi-) deterministic world among the Universe is ignorable, but based on the ontological perspective of the MWI, such worlds exist actually and are as real as ours as we perceive and cognize it. Thus, despite the ontologies of (semi-) determinist agents are proposed less frequently, according to the argument of enquiry \textbf{Q2} they release true pictures of the possible universe. Therefore, it seems that the chance of detection would not resolve the discovered paradox properly.

\par Also one may contemplate about a proposal for \emph{relativity in ontology} of parallel worlds, assuming no conflict in disparity of ontologies via transferring from a parallel world to another one. That is, we should trust in all possible universes proposed by the agents of the parallel worlds in the MWI as relative ontologies. However, if we try to rely on this proposal, we must in turn admit more complete universes with respect to what we consider in the MWI, and hence embrace any illusive theory of nature (illusive world) which is unreal based on the objective reality we cognize in our world. For instance, we can assume the Pauli's exclusion principle as a deterministic law imposed to our world among a larger possible universe to turn us to semi-determinist agents for a more extended theory of quantum mechanics. Thus, since the mentioned theory does not include the Pauli's exclusion principle, there would be some real parallel worlds with no definite structures of atoms and no acknowledged Mendeleev's table. In principle, pursuing this proposal would lead us to uncontrollable theories of exotic real worlds, an endless diverging illlusive process. 

\par In fact, the problems of the MWI has some more complicated features through with the picture we presented here. As we argued, since in the MWI the existence of any possible world is considered as a real fact, the established laws of nature of any world may probably change from a definite moment onward (after splitting from a branching point). Actually, this is the most troubling situation when we assume the laws of physics of our world be suddenly broken and replaced with some different laws of nature.\footnote{Also, such a catastrophic phenomenon would have happened in the past, and then we would likely be encountered with a pseudoscientific theory of physics which can confirm (if we insist more inquisitively) the mythological stories.} All in all, the MWI results in the fact that the recorded reports of observable reality could be based on transient objective laws and unstable reality. Consequently, there would be no guarantee that quantum mechanics become suddenly extinct and be replaced with a new theory of nature. Particularly, it must be a catastrophic end of physics. Thus, despite of its interesting expressions of peculiar aspects of quantum mechanics, the MWI seems not to be a correct interpretation of the theory. Therefore, although the Copenhagen interpretation of quantum mechanics has been established to be inconsistent, it cannot be safely replaced with the MWI.





\section{Acknowledgments}

\par The author says his special gratitude to S. Ziaee. Moreover, this article is dedicated to medical staff from all around the world who are selflessly fighting the pandemic of Covid-19. Also it should be noted that this research was in part supported by a grant from IPM (No.99810422).


\end{document}